# Instabilities in the vortex matter and the peak effect phenomenon

Shyam Mohan, Jaivardhan Sinha, and S. S. Banerjee[a]

Department of Physics, Indian Institute of Technology, Kanpur-208016, Uttar Pradesh,

India

Yuri Myasoedov

Department of Condensed Matter Physics, Weizmann Institute of Science, Rehovot

76100, Israel

## Abstract

In single crystals of 2H-NbSe$_2$, we identify for the first time a crossover from weak collective to strong pinning regime in the vortex state which is not associated with the peak effect phenomenon. Instead, we find the crossover is associated with anomalous history dependent magnetization response. In the field ($B_{dc}$) - temperature ($T$) vortex matter phase diagram we demarcate this pinning crossover boundary. We also delineate another boundary which separates the strong pinning region from a thermal fluctuation dominated regime, and find that PE appears on this boundary.

PACS numbers: 74.25.Qt, 64.70.Pf, 74.25.Dw, 74.70.B

---

[a] email: satyajit@iitk.ac.in

The statics and dynamics of elastic media in a random pinning environment is common to a variety of systems like the vortex state in superconductors[1], Wigner crystals[2], charge density waves[3], magnetic domains[4] etc. The elastic vortex matter experiences a perennial tussle between elastic forces trying to order it and thermal fluctuations and pinning trying to disorder it, leading to a variety of phenomenon. Two widely observed phenomena are, the thermally driven first order melting[1] of the vortex lattice in high $T_c$ superconductors and the peak effect (PE)[5] in low $T_c$ superconductors. While both phenomena are related to disordering of the vortex lattice, the PE phenomenon still lacks a comprehensive understanding. The ubiquitous PE phenomenon widely observed in a large variety of superconductors is an anomalously large enhancement in the critical current density ($J_c$) (or equivalently the pinning force density $= \vec{J}_c \times \vec{B}_{dc}$) close to the superconducting – normal boundary. While it is known that the vortex configuration changes across the PE viz., the ordered vortex lattice disorders[6], one is yet to be certain about the mechanism which triggers the rise in pinning across the PE. To date no theory has been able to quantitatively explain the extent of rise in $J_c$ across the PE. Many theories[7,8,9,10] have investigated the effects of pinning (both weak and strong) on the configuration of vortices in isotropic and anisotropic superconductors[11]. The vortex configuration in a sample of a type II superconductor with strong intrinsic pinning is more disordered[9] than a relatively well ordered configuration[10] in a sample with weaker pinning. However, in a sample with both weak and strong pins, how does the vortex matter behave? The competition between elasticity of the vortex lattice and random pinning present in the sample, governs which interaction wins in the vortex matter. To understand the mechanism of PE, it maybe relevant to look at the different pinning regimes possible in the vortex matter. In literature two major pinning regimes for the vortex state in superconductors have been identified. On the one hand the collective pinning theory (CP)[7,8] describes the collective action of '*weak pins*' and on the other the flux pinning theory by Labusch[12] describes the independent action of '*strong pinning*' centers on vortices. The effective pinning force in the CP theory is determined by calculating the extent of short range order present in the vortex matter created by weak pins trying to distort the rigid elastic medium of the vortex lattice. The Labusch theory on the other hand determines the pinning force from the competition of strong pins trying to distort an elastic vortex line. A recent theory[13] argues that the PE phenomenon occurs naturally due to an increase in $F_p$ (and hence in $J_c$) associated with a crossover from the weak collective to a strong pinning regime. In the present paper based on magnetization measurements we have identified the weak to strong pinning crossover regime. We find that this crossover is associated with a small change in $J_c$, which is *insufficient* to produce the PE. Rather than being associated with the crossover in pinning, we find that the PE phenomenon is situated in a special region of the phase diagram, which is on a boundary separating the strong pinning regime from a region dominated by thermal fluctuations. The close proximity of PE to a region dominated by thermal fluctuations implies that the phenomenon is a complex superposition of both pinning and fluctuation effects.

Ac and dc magnetization measurements have been performed in two single crystals of 2H-NbSe$_2$ (labeled as #1 and #2). The two crystals #1 and #2, have similar average dimensions of *1.5×1.5×0.1 mm³* and *T$_c$(0) ≈ 7.2 K* and *T$_c$(0) ≈ 7.1 K* respectively. To search for the weak to strong pinning crossover we selected single crystals with very weak pinning (i.e., *J$_c$ ~ 500 A/cm²* at *5.0 K)*. To reduce the possibility of strong pinning generated by extended defects likely to be present along the c-axis in layered 2H-NbSe$_2$ and also, to avoid geometric and surface barrier effects which persist upto the PE in *B$_{dc}$ // c* orientation[14], we have chosen the *B$_{dc}$ // ab* direction for our measurements. Due to the layered nature of 2H-NbSe$_2$, conventional transport measurements are difficult in *B$_{dc}$ // ab* orientation. To obtain information about the pinning in the static state of the vortex matter we investigate the ac susceptibility and dc magnetization response of the sample using a commercial Quantum Design SQUID magnetometer (model no. MPMS-XL5) and an Oxford VSM (model no. 3001).

The ac susceptibility response ($\chi'$ and $\chi''$) of the sample was recorded as a function of temperature (T) at fixed dc magnetic fields (*B$_{dc}$*) for different values of the ac drive (*h$_{ac}$*) at a frequency of *211 Hz*, on the SQUID magnetometer using the Reciprocating Sample Option (RSO) for increased sensitivity and to reduce field inhomogeneity effects on the magnetization[b]. Shown in Fig.1(a) is the $\chi'(T)$ response for *B$_{dc}$ = 100 G* for different *h$_{ac}$*. Anomalous enhancement in pinning associated with the peak effect (PE) causes the sample to shield the penetrating $h_{ac}$ more efficiently from within its interior, thereby enhancing the diamagnetic ($\chi'$) response at the onset of PE, which begins at around *7.04 K* in Fig.1 (a). For interpreting our results on $\chi''$, it would be worthwhile to recall that the out of phase ac susceptibility, $\chi''$, is a measure of the dissipation[15] caused by the dragging of *normal* vortex cores which are oscillating under the influence of the periodically varying *ac* magnetic field ($h_{ac}$). Insufficient penetration of $h_{ac}$ into the sample or an enhancement in the pinning of vortices, would lower the vortex dissipation which corresponds to a low $\chi''$ response. From Fig. 1(b) it is clear that for $h_{ac} < 1\ G$ and at low *T*, due to almost complete shielding of the probing ac magnetic field from the bulk of the sample, the $\chi''(T)$ response is nearly zero. Fig. 1(b) shows that at fixed *T* (at say *T = 6.7 K*) as $h_{ac}$ increases, the $\chi''$ response also increases monotonically. Full penetration of $h_{ac}$ up to the sample center causes a significant rise in dissipation in the sample, which in turn leads to a broad maximum in the $\chi''$ response[16] (location marked as A in Fig.1(b) for $h_{ac} = 2\ G$). At the PE region due to enhancement in pinning we observe a drop in the dissipation ($\chi''$) response (marked as $T_p$ in Fig.1 (b)). Beyond $T_p$, dissipation has a tendency to rise sharply before decreasing close to $T_c(H)$ (we discuss this feature in the next section). In the absence of PE, if the pinning in the vortex state did not change, then beyond the broad maxima at A in Fig.1(b) the $\chi''$ response should crossover smoothly into the enhanced dissipation regime (viz., the region beyond $T_p$ in Fig.1(b)) close to $T_c(H)$.

---

[b] See writeup on RSO option from Quantum Design in the applications notes at the following web address
http://www.qdusa.com/resources/pdf/mpmsappnotes/1014-820.pdf

The main panel of Fig. 2 shows the behavior of $\chi''(T)$ response at different $B_{dc}$ (> 750 G). From $\chi'(T)$ measurements we find PE phenomenon disappears above *750 G* in our crystals. For $B_{dc} = 12500\ G$, we have identified three distinct regimes of behavior in the $\chi''(T)$ response. In region 1, the high dissipation response emanates from full penetration of $h_{ac}$ to the center of the sample[16], similar to the response at A in Fig.1 (b). One would have expected that in the absence of the PE phenomenon, the region 1 should smoothly have joined into a high dissipation region close to $T_c(H)$ (similar to region above $T_p$ in Fig.1(b)). Instead, in the cross shaded region 2 we see a new behavior in the dissipation ($\chi''$) response, viz., in this region located between the two arrows there is a substantial decrease in the dissipation. Inset (a) of Fig. 2, shows the onset of the drop in dissipation (marked as $T_{cr}$) determined from the derivative, $d\chi''/dT$. From $d\chi''/dT$ and the main panel of Fig.2, we see that subsequent to the drop in $\chi''(T)$ in region 2, the dissipation response attempts to show an abrupt increase at the onset of region 3 (marked as $T_{fl}$ in the inset (a) and main panel of Fig.2). The abrupt increase in dissipation beyond $T_{fl}$ is more pronounced at low $B$ and high $T$. For $B_{dc} < 750\ G$, the $T_{fl}$ location is the same as $T_p$ (see Fig.1(b) where dissipation enhances above $T_p = T_{fl}$). The inset (b) of Fig.2 shows the absence of PE at $T_{cr}$ in the $\chi'(T)$ response at 1000 G and 12500 G, indicating that the anomalous drop in dissipation in region 2 is not associated with the PE phenomenon. An alternative way of investigating the nature of pinning above $T_{cr}$ is by quenching the vortex state (field cooling (FC)) from $T > T_{cr}$. Our observation of a low dissipation ($\chi''$) response in the FC state (cf. main panel of Fig.2 at 1000 G), implies that the pinning enhances across $T_{cr}$. Above $T_{cr}$ the high pinning regime exists till $T_{fl}$.

The tendency of the dissipation to rapidly rise close to $T_{fl}(B)$ is a behavior which is expected across the irreversibility line ($T_{irr}$), where the bulk pinning in the superconductors vanishes. We have confirmed that $T_{fl}(B)$ coincides with $T_{irr}(B)$, by comparing dc magnetization with $\chi''$ response measurements (cf. arrow marked as $T_{fl} = T_{irr}$ in the main panel of Fig.3).

Fig.3 shows the magnetization hysteresis in the two crystals of 2H-NbSe$_2$ measured on a SQUID and VSM. The inset (a) of Fig.3 shows the M-H hysteresis loop recorded at *6K*. A striking feature of the M-H loop is the asymmetry in the forward and reverse legs of the magnetization hysteresis response. A feature which can easily be missed on the scale of the full magnetization hysteresis loop is the small change in curvature (marked with an arrow in inset (a) of Fig.3) on the $M_{rev}$ leg of the hysteresis loop. In the main panel of Fig. 3, we have plotted only the $M_{rev}$ recorded at different temperatures. At the locations marked with arrows in Fig.3 there is a substantial change in slopes of the $M_{rev}$ curves. The characteristic bump like feature (marked with the arrow) is observed at different $T$ and only on the $M_{rev}$ curve but not on the $M_{for}$ leg. This strong history dependence is in a region of the $B$ - $T$ phase diagram, which is far from the PE region. We have confirmed all the above new features in ac and dc magnetization measurements in another sample of 2H-NbSe$_2$ (#2) with similar weak pinning in the $B_{dc}$ // $ab$ orientation.

Fig. 4 shows the $B_{dc}$ - $T$, vortex matter phase diagram wherein we show the location of the $T_c(B)$ line which is determined by the onset of diamagnetism in $\chi'(T)$, the $T_p(B)$ line which denotes the location of the PE phenomenon, the $T_{cr}(B)$ line across which the $\chi''(T)$ response (shaded region 2 in Fig.2) shows a substantial decrease in the dissipation and the $T_{fl}$ line beyond which dissipation attempts to increase. The PE ceases to be a distinct noticeable feature beyond *750G* and the $T_p$ line continues as the $T_{fl}$ line. Unlike the behavior[17] of the $T_p(B)$ line, which usually runs parallel to $T_c(B)$, the $T_{cr}(B)$ line has a distinct curvature that extrapolates to zero close to $T_c(0) \sim 7.2$ K.

We consider the $T_{cr}(B)$ line as a crossover in the pinning strength experienced by vortices, which occurs well prior to the PE. A criterion[1,13] for weak to strong pinning crossover is: the change in the pinning force far exceeds the change in the elastic energy of the vortex lattice, due to pinning induced distortions of the vortex line. This can be expressed as[13], the pinning force $(f_p)$ ~ Labusch force $(f_{Lab}) = (\varepsilon_0\xi/a_0)$, where $\varepsilon_0 = (\phi_0/4\pi\lambda)^2$ is the energy scale for the vortex line tension, $\xi$ is the coherence length, $\phi_0$ flux quantum associated with a vortex, $\lambda$ is the penetration depth and $a_0$ is the inter vortex spacing ($a_0 \propto B^{-0.5}$). A softening of the vortex lattice satisfies the criterion for the crossover in pinning. At the crossover in pinning we have a relationship, $a_0 \approx \varepsilon_0\xi f_p^{-1}$. At $B=B_{cr}$ and far away from $T_c$, if we use a monotonically decreasing temperature dependent function for $f_p \sim f_{po}(1-t)^\beta$, where $t=T/T_c(0)$ and $\beta > 0$, then we obtain the relation $B_{cr}(T) \propto (1-t)^{2\beta}$. We have used the form derived for $B_{cr}(T)$ to obtain a good fit (red dotted line in Fig.4) for $T_{cr}(B)$ data, giving $2\beta \sim 1.66 \pm 0.03$. Inset of Fig. 4 is a log-log plot of the width of the magnetization loop ($\Delta M$) versus $B_{dc}$. Upon reducing the magnetic field from the upper critical field, $\Delta M \propto J_c$ (or equivalently pinning) increases upto $B_{cr}$. $\Delta M$ subsequently decreases at $B < B_{cr}$ and smoothly crosses over to a weak collectively pinning regime. The weak collective pinning regime[18] is characterized by the region shown in the inset, where the measured $\Delta M(B)$ (red) coincide with the black dashed line viz., $\Delta M \propto J_c \propto 1/B_{dc}^p$ with $p$ a positive integer,.

The shaded region (in green) in the $\Delta M(B)$ plot shows the excess pinning that develops due to the pinning crossover across $B_{cr}$. The distinctness of the $T_{cr}$ and $T_p$ lines in Fig.4 shows that the excess pinning associated with the pinning crossover does not produce any PE. Based on the above discussion we surmise that the $T_{cr}(B)$ line marks the onset of an instability in the static vortex lattice due to which there is a crossover from weak (region 1 in Fig.2 main panel) to a strong pinning regime (region 2 in Fig.2 main panel). The crossover in pinning produces interesting history dependent response in the superconductor, as seen in the $M_{rev}$ measurements of Fig. 3 and in the $\chi''(T)$ response for the zero field cooled response (ZFC) and FC vortex states, in the main panel of Fig.2. In the inset (b) of Fig.3 we have schematically identified the pinning crossover by distinguishing two different branches in the $M_{rev}(B)$ curve, which correspond to magnetization response of vortex states with high and low $J_c$. The reasons for the instability across $T_{cr}(B)$ could be a softening of the elastic modulii of the lattice due to the proliferation of topological defects in the static vortex lattice[10]. It is interesting to note that a similar behavior has been observed in the driven vortex state, as deduced from transport measurements[19]. We reiterate that the onset of instability or weakening of the elastic modulii of the lattice sets in well prior to PE phenomenon without producing the anomalous PE.

As the strong pinning regime commences upon crossing $B_{cr}$, how then does pinning dramatically enhance across PE? The $T_{fl}(B)$ line in Fig.4 marks the end of the strong pinning regime of the vortex state. Above the $T_{fl}(B)$ line and close to $T_c(B)$, the tendency of the dissipation response to increase rapidly (Figs.1 and 2) especially at low B and high T, implies that thermal fluctuation effects dominate over pinning. We find that our values ($B_{fl}$, $T_{fl}$) in Fig.4, satisfies the equation governing the melting of the vortex state[1], viz., $B_{fl} = \beta_m \left(\frac{c_L^4}{G_i}\right) H_{c2}(0) \left(\frac{T_c}{T_{fl}}\right)^2 \left[1 - \frac{T_{fl}}{T_c} - \frac{B_{fl}}{H_{c2}(0)}\right]^2$, where, $\beta_m = 5.6$ (Ref.1), Lindemann no. $c_L \sim 0.25$ (Ref. 6, A. M. Troyanovski et al.), $H_{c2}^{//ab}(0) = 14.5$ T, if a parameter, $G_i$ is in the range of $1.5 \times 10^{-3}$ to $10^{-4}$. The Ginzburg number, $G_i$, in the above equation controls the size of the $B_{dc}$ - $T$ region in which thermal fluctuations dominate. A value of $O(10^{-4})$ is expected for 2H-NbSe$_2$ (Ref.17, M. J. Higgins et al..). The above discussion implies that thermal fluctuations dominate beyond $T_{fl}(B)$. By noting that $T_p(B)$ appears very close to $T_{fl}(B)$ it seems that PE appears on the boundary separating strong pinning and thermal fluctuation dominated regimes.

To surmise, we have found evidence in dissipation ($\chi''$) measurements for a weak to strong pinning crossover in very weakly pinned crystals of 2H-NbSe$_2$. We have found that the pinning crossover is located far from the PE region. The pinning crossover is also associated with interesting history dependent magnetization response. Our observations imply that instabilities developing within the vortex lattice (perhaps a softening of the elastic modulii) leads to the crossover in pinning which occurs well before the PE. In fact, PE seems to sit on a boundary which separates a strong pinning dominated regime from a thermal fluctuation dominated regime. Our assertion has significant ramifications pertaining to the origin of PE which was originally attributed to a softening of the elastic modulii of the vortex lattice[5]. Even though thermal fluctuations try to reduce pinning, we believe our results show that through PE the two combine in a non trivial way to enhance the pinning dramatically. The close proximity of PE to the thermal fluctuation dominated regime implies that perhaps, one also needs to investigate the nature of the superconducting order parameter in the PE region. We speculate that PE maybe be influenced by effects related to a weakening of the superconducting order parameter close to $T_c(B)$, making the superconductor more susceptible to local perturbations close to this region. We hope our results would pave the way for a fresh approach towards understanding the origins of the puzzling phenomenon of PE.

The authors would like to acknowledge the support from A. K. Grover, D. Chowdhury and Eli Zeldov. R. Sharma is thanked for technical assistance. S. S. B. gratefully acknowledges a research grant from Prof. S. G. Dhande, Director, IIT Kanpur.

---

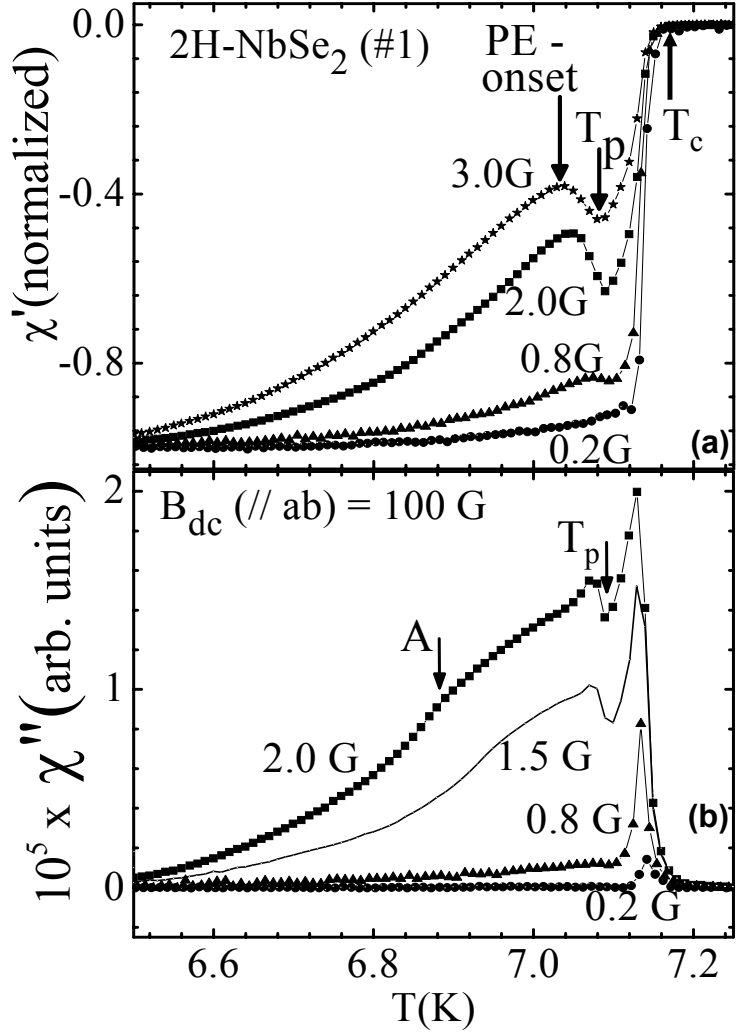

**Figure.1**: The panels (a) and (b) show the amplitude dependence of the $\chi'(T)$ and $\chi''(T)$ response measured for the sample 2H-NbSe$_2$ (#1) for $B_{dc} = 100\ G\ (//\ ab)$ at different $h_{ac}$, with $f = 211\ Hz$.

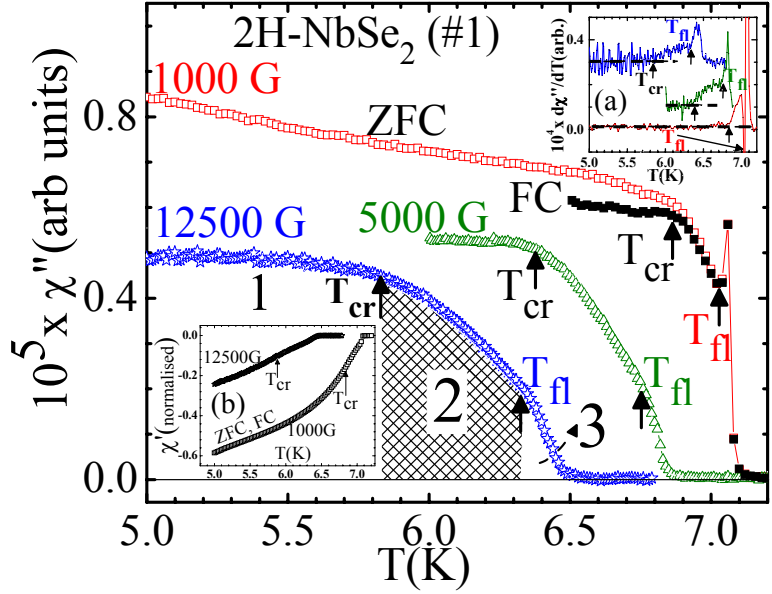

**Figure.2**: Indicates the anomalous dissipation regime (marked as the shaded region 2) in the $\chi''(T)$ response measured at $h_{ac} = 2G$ and $f = 211\ Hz$ at different $B_{dc}$. The onset of the anomalous dissipation regime is marked as $T_{cr}$. Inset (a) shows the location of $T_{cr}$ and $T_{fl}$ in $d\chi''/dT$ for 1000 G (red), 5000 G (green) and 12500 G (blue). Inset (b) shows the absence of PE at $T_{cr}$ in the $\chi'(T)$ response. The main panel also shows the difference between the ZFC and FC $\chi''(T)$ response at 1000 G below $T_{cr}$, which is absent in the $\chi'(T)$ response.

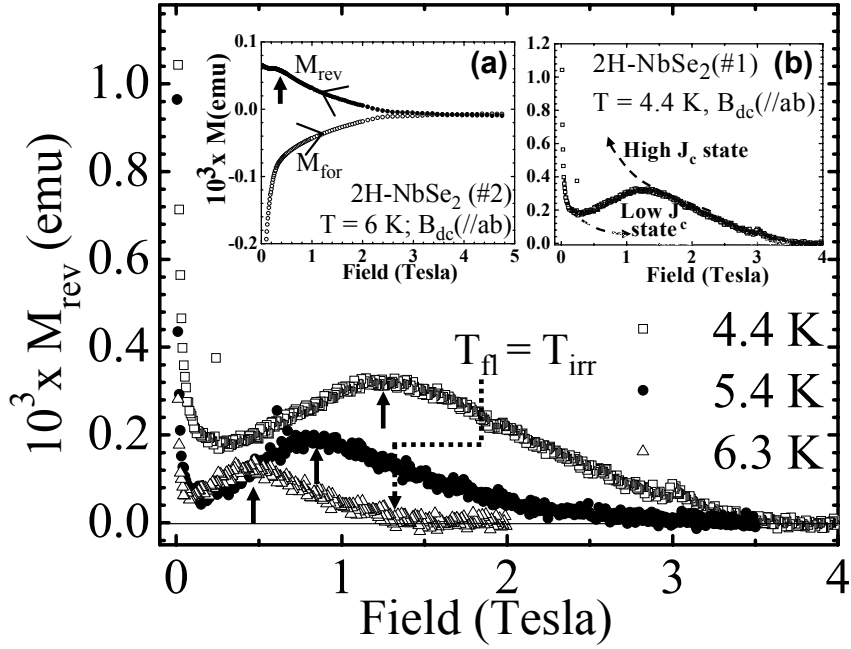

**Figure.3**: Main panel shows the $M_{rev}(B)$ portion of the magnetization hysteresis loop measured for the 2H-NbSe$_2$ (#1) sample at different T. Inset (a) demonstrates the asymmetry of the magnetization hysteresis loop. The arrow in the inset (b) shows the 'bump' in $M_{rev}(B)$. The figure also schematically illustrates the two branches of magnetization response which correspond to vortex states with high and low $J_c$ (or pinning).

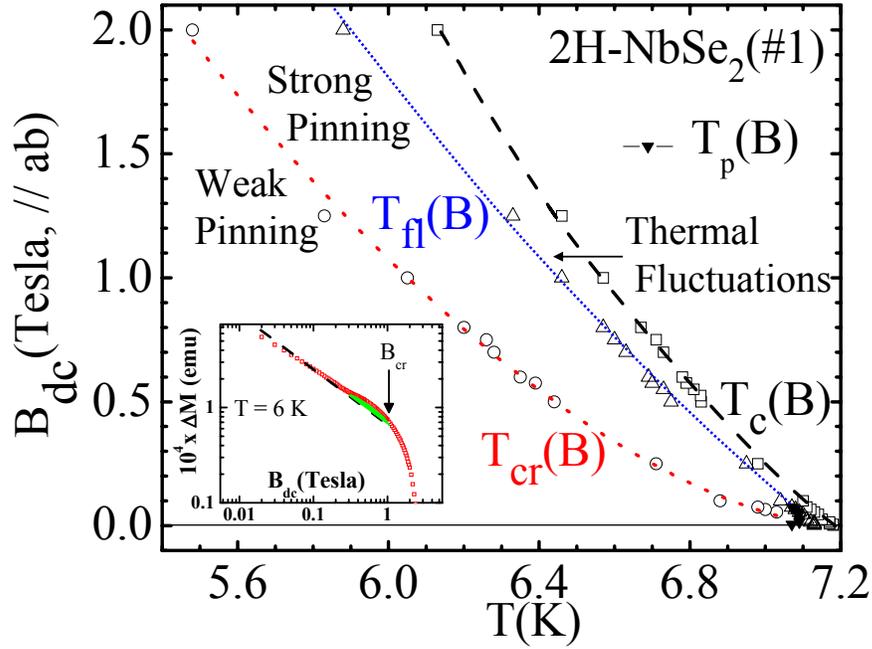

**Figure.4**: The main panel shows the $T_{cr}(B)$, $T_p(B)$, $T_{fl}(B)$ and $T_c(B)$ boundaries in the phase diagram. The dotted line through $T_{cr}(B)$ data is a fitted line (refer to text for details). The inset is a *log-log plot of ΔM* vs. $B_{dc}$. In the inset, the red data points coinciding with the black dashed line indicates the weak collectively pinning regime of the vortex matter.